\documentclass{article}
\usepackage{ijcai23}

\usepackage{times}
\usepackage{soul}
\usepackage{url}
\usepackage[hidelinks]{hyperref}
\usepackage[utf8]{inputenc}
\usepackage[small]{caption}
\usepackage{graphicx}
\usepackage{amsmath}
\usepackage{amsthm}
\usepackage{booktabs}
\usepackage{algorithm}
\usepackage{algorithmic}
\usepackage[switch]{lineno}

\usepackage{xcolor}

\usepackage{mathtools}
\usepackage{siunitx}

\usepackage{xspace}

\title{The \Acc: Combining Reactivity, Robustness, and Musical Expressivity in an Automatic Piano Accompanist}

\iftrue
\author{
Carlos Cancino-Chacón$^1$
\and
Silvan Peter$^1$\and
Patricia Hu$^1$\and
Emmanouil Karystinaios$^1$\and\\
Florian Henkel$^2$\and
Francesco Foscarin$^1$\and
Nimrod Varga$^1$\And
Gerhard Widmer$^1$
\affiliations
$^1$Institute of Computational Perception, Johannes Kepler University Linz, Austria\\
$^2$SiriusXM + Pandora, USA\\
\emails
\{carlos\_eduardo.cancino\_chacon, silvan.peter, patricia.hu\}@jku.at
}
\fi

\begin{document}
\newcommand{\Acc}{ACCompanion\xspace}
\newcommand{\SolScore}{\ensuremath{Z}}
\newcommand{\SolPerf}{\ensuremath{X}}
\newcommand{\AccScore}{\ensuremath{W}}
\newcommand{\AccPerf}{\ensuremath{Y}}

\newcommand{\SolScoreN}{\ensuremath{z}}
\newcommand{\SolPerfN}{\ensuremath{x}}
\newcommand{\AccScoreN}{\ensuremath{w}}
\newcommand{\AccPerfN}{\ensuremath{y}}

\maketitle

\begin{abstract}
    This paper introduces the \Acc, an expressive accompaniment system. 
    Similarly to a musician who accompanies a soloist playing a given musical piece,
    our system can produce a human-like rendition of the accompaniment part that follows the soloist's choices in terms of tempo, dynamics, and articulation. 
    The \Acc works in the symbolic domain, i.e., it needs a musical instrument capable of producing and playing MIDI data, with explicitly encoded onset, offset, and pitch for each played note.  
    We describe the components that go into such a system,
    from real-time score following and prediction to expressive performance generation and online adaptation to the expressive choices of the human player. 
    Based on our experience with repeated live demonstrations in front of various audiences, we offer an analysis of the challenges of combining these components into a system that is highly reactive and precise, while still a reliable musical partner, robust to possible performance errors and responsive to expressive variations. 
\end{abstract}

\section{Introduction}

\begin{figure}
    \centering
    \includegraphics[width=\columnwidth]{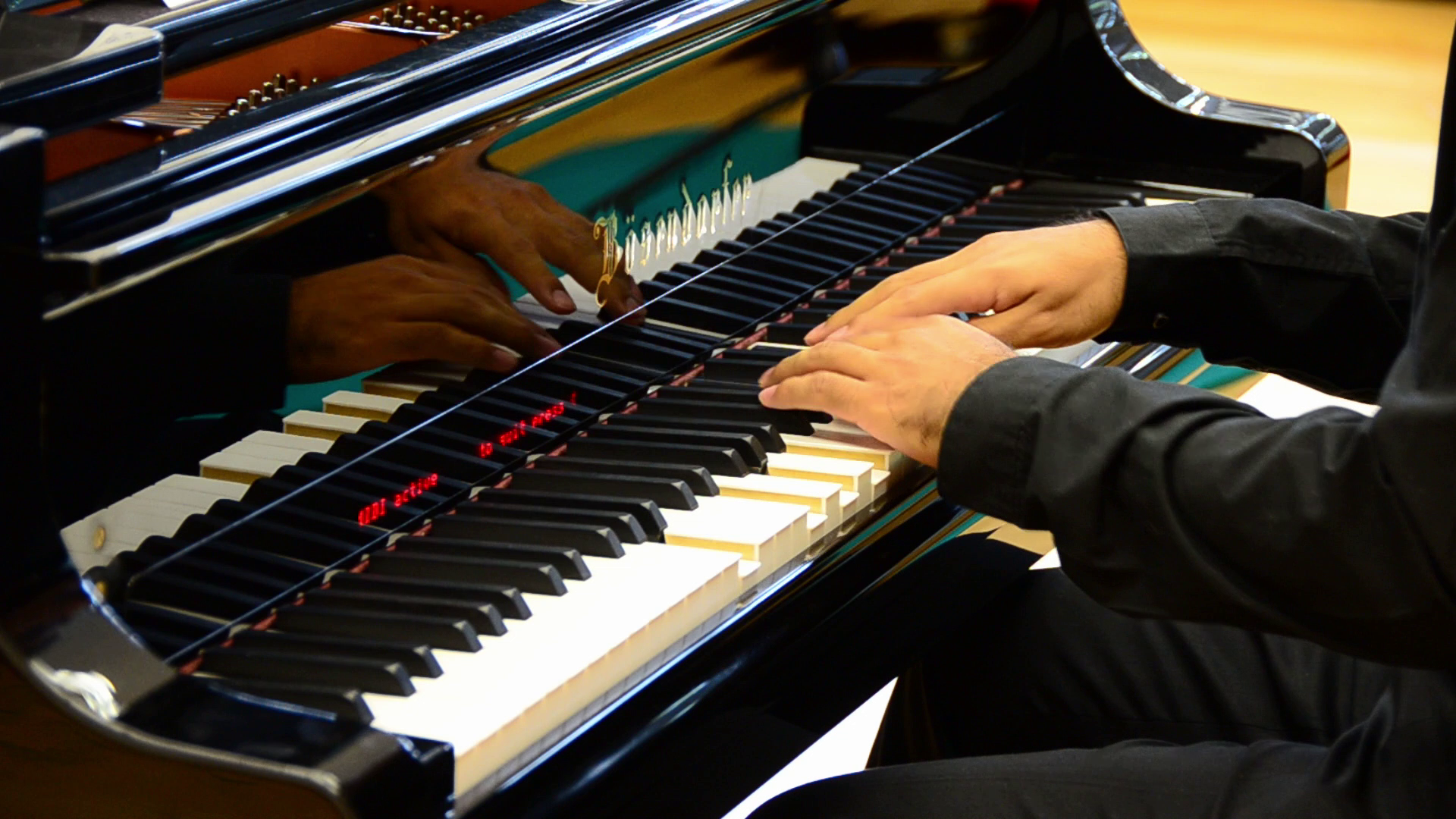}
    \caption{A photo of the \Acc in action using a computer-controlled piano. Some notes are played by the human soloist, and some by our system.}
    \label{fig:accompanion_poster}
\end{figure}

Participating in ensemble music playing requires creative collaboration and can provide fulfilling musical experiences. Given their success in other creative domains, there is an emerging trend towards the exploration of AI systems in interactive and collaborative settings.
Interactive accompaniment systems are computer programs that can perform jointly with human musicians. In such human-machine musical collaboration settings, the computer system can either improvise its musical content or render an expressive accompaniment according to a given score in response to the human player in real-time~\cite{dannenberg1987following,dannenberg1984line,vercoe1984synthetic}.
We target the latter case and we restrict our attention to the symbolic domain, i.e., we work with music data that, as opposed to audio files, explicitly encode high-level note features such as pitch, onset (when a note starts), and duration.
Specifically, we propose the \Acc, an automated accompaniment system that, given a symbolically encoded score with a soloist and an accompaniment part, is able to render an expressive MIDI performance of the accompaniment in response to the MIDI performance of a human soloist.
This system can work in different configurations, as long as the soloist performances can be encoded in MIDI data, and a MIDI player that translates MIDI outputs to sound is available~(see a photo of the system in action on a computer-controlled player piano in Figure~\ref{fig:accompanion_poster}).

Since the development of the first MIDI accompaniment systems~\cite{dannenberg1984line,vercoe1984synthetic}, most research in this field has focused on the synchronization aspect, that is, matching the accompaniment with the human performance~\cite{nakamura2015real,chen2014improved,Raphael2009OrchestralAF}. In terms of musical expression in the accompaniment performance, most systems to date constrain their expressive response to individual expressive aspects such as tempo~\cite{cont2008antescofo,raphael2010music}, or dynamics and timing~\cite{xia2015spectral}. We address the issue of musical expression in the accompaniment performance by explicitly modeling tempo, dynamics and articulation aspects into the \Acc, making it capable of generating expressive accompaniments in response to the performance style of the soloist.
We also compare cognitive plausible tempo models inspired by the research in the sensorimotor synchronization.

Moreover, we have showcased our system in a number of real-world scenarios, including 
demonstrations to the scientific and general public (e.g., at the Falling Walls Science Summit 2021 in Berlin, at Gerhard Widmer's Keynote at IJCAI-22 in Vienna and at the Heidelberg Laureate Forum Foundation). The \Acc won an Award for Creative Achievement at the AccompaniX competition organized by the Neukom Institute for Computational Science at Dartmouth College in 2017.
However, our own experience also made clear that although our system can perform interactively and expressively with a human soloist, ensemble music performance and shared musical interpretations between humans involve a level of cognitive-emotional alignment that cannot yet be achieved.
Given our experience, we want to start a critical discussion on the causes and possible solutions to this problem. Finally the \Acc will also be published as open source software.\footnote{Supplementary materials and code for this paper can be found at the following link: \url{https://cpjku.github.io/accompanion\_ijcai2023/}} 

Overall, our contributions in the context of automatic accompaniment systems are as follows: (1) the generation of an accompaniment part that is conditioned on the soloist performance on three expressive parameters, that is, tempo, dynamic, and articulation; (2) the exploration of cognitively plausible tempo models, inspired by research from the field of sensorimotor synchronization; (3) an extensive system evaluation in various real-world scenarios, including concerts in public venues and critical exposure to professionally trained musicians.
The rest of this paper is organized as follows: 
Section~\ref{sec:rel_work} presents related work and Section~\ref{sec:architecture} outlines the \Acc system. 
The key components are evaluated in Section~\ref{sec:evaluation}, followed by a human performer's perspective and a discussion on collaborative human-machine musical modeling in Section~\ref{sec:discussion}. 
Conclusions and future research recommendations are presented in Section~\ref{sec:conclusion}.

\section{Related Work}\label{sec:rel_work}

Dannenberg~\shortcite{dannenberg1984line} identifies three tasks that accompaniment systems must perform to play effectively with a human:
\begin{enumerate}
\item \emph{Soloist part detection}: capturing a human performance in real-time and identifying the performed notes from its representation.
\item \emph{Score following}: precisely matching these performed notes to notes in the score, while being robust against possible player errors. 
\item \emph{Expressive accompaniment generation}: producing an expressive accompaniment part that is conditioned on the soloist part.
\end{enumerate}

The detection step is a prominent part of systems that work from audio recordings. Since we use MIDI data, this step is handled by the MIDI instrument, which explicitly encodes the note information.
In this section, we present related work on the score following and expressive accompaniment tasks, as well as some relevant perspectives from the music cognition literature on joint music performance.

\subsection{Score Following}\label{sec:rel_score_foll}
The score following task can be performed in an online or offline setting, starting from MIDI or audio. 
In an online, MIDI-based context, we aim to perform \textit{online symbolic score following}, aligning each note in a MIDI performance with corresponding elements (time or notes) in the musical score. 
Recent work by Raphael and Gu~\shortcite{Raphael2009OrchestralAF} and Nakamura et al.~\shortcite{Nakamura2014MergedOutputHM}, both employ Dynamic Bayesian Networks for this task. 
Our \Acc employs a hidden Markov model (HMM) based score follower that is roughly comparable to the one by Raphael and Guo in design (see Section~\ref{sec:score_follower}).

Related work has been developed in the domain of online audio-to-score alignment that aligns every temporal position in the performance with a temporal position in the score~\cite{dixon2005line,cont2008antescofo,raphael2010music,duan2011state,Arzt2015RealTimeMT}. 
These systems are based on On-Line Time Warping (OLTW) or variants of HMMs.
A similar range of approaches, albeit in a non-causal formulation, is used for offline symbolic score following~\cite{gingras2011improved,chen2014improved,nakamura2017performance}. Recently, Peter et al.~\shortcite{peter2023tismir} performed a systematic evaluation of dynamic time warping (DTW) versus HMM-based approaches for offline symbolic alignment, and found that they yield similar results for the alignment of piano classical pieces. 
DTW-based approaches have therefore the potential to be a viable (but still untested) alternative to HMMs for online symbolic score following, and they could have a different set of strengths and weaknesses. For this reason, in addition to the HMM mentioned above, the \Acc also contains an OLTW score follower, which adapts to the symbolic domain many ideas from the state-of-the-art in audio score following~\cite{Arzt2015RealTimeMT}.
Note that the score follower is only a part of our system, and it is out of the scope of this paper to perform a systematic evaluation of the state-of-the-art in this field.

Gingras and McAdams~\shortcite{gingras2011improved} use a tempo model to ``smooth'' erratic score follower outputs due to embellishments or player mistakes. 
Similarly, Raphael and Gu~\shortcite{Raphael2009OrchestralAF} use a Kalman filter for this purpose and for predicting future note positions. 
We explore multiple variations of such a tempo model. 
Other recent systems~\cite{arzt2018audio,agrawal2020hybrid} suggest employing deep learning for enhanced audio feature pre-processing prior to alignment.

\subsection{Expressive Accompaniment Generation}
For generating an expressive accompaniment, a system needs to solve two problems.
The first is to generate an expressive rendition of a musical piece that conforms to human understanding and perception of music and communicates its inherent emotional and affective content. The second consists of adapting this expressive rendition in response to the performance style of the soloist.
Some work in the literature focus on the first problem alone with rule-based algorithms~\cite{friberg2006overview}, probabilistic approaches \cite{widmer2009yqx}, or, recently, neural networks \cite{jeong2019virtuosonet,fujishima2018rendering,cancino2018computational}.
A detailed description of these models falls outside the scope of the current paper; the interested reader is directed to \cite{cancino2018computational}. 
Note that all these systems work offline and are not suitable for a real-time scenario. However, they can be used for a first offline generation step, which gets then adapted to solve the second problem, i.e., to condition the generated accompaniment on the soloist performance.

Multiple works target this second problem but focus only on certain expressive performance parameters.
Cont ~\shortcite{cont2008antescofo} explicitly models the expressive tempo to enable the temporal interaction between the human performer and the accompaniment system, while not considering other expressive aspects, such as dynamics or articulation. 
Raphael~\shortcite{raphael2010music} uses a Gaussian graphical model as a tempo model to schedule the accompaniment performance, but otherwise constructs the performance itself by time-stretching a prerecorded audio recording of the accompaniment.
More recently, Xia et al. \shortcite{xia2015spectral,xia2015duet} proposed to model a duet ensemble as a linear dynamic system with learned parameters. At each position in the score, the time and dynamics of the next accompaniment notes are predicted as linear combinations of some local note features, ``smoothed'' by a hidden variable that acts similarly to the tempo model of Raphael. 

\subsection{Perspectives from Music Cognition}\label{sec:sms}
From a cognitive standpoint, we can consider an ensemble performance as a social event in which individual musicians are engaged in a joint action that results in feelings of musical togetherness. Such a joint action can be described in terms of the underlying sensorimotor synchronization (SMS), that is, the processes involved in the temporal coordination of an action or movement with an external rhythm~\cite{repp2006musical}. 

Mathematically, SMS can be modelled via either the information-processing approach or via dynamical systems theory \cite{loehr2011temporal}.
We follow the former, which assumes the existence of a timekeeper instance to count successive actions and generate motor commands in response to external stimuli. 
In the context of ensemble music performances, the rhythmic sensory stimuli are usually characterized by irregularities (expressive timing), resulting in some degree of uncertainty with regard to the precise timing of the next event. 
For the timekeeper to achieve sustained SMS despite this variability, some form of error correction and/or temporal prediction is necessary \cite{vorberg1996modeling,mates1994model,van2013adaptation}. We consider these findings in our modeling choices for the underlying tempo models, which we describe in Section~\ref{sec:accompanist} and evaluate in Section~\ref{sec:tempo_models}.

\section{System Architecture}\label{sec:architecture}

\begin{figure}
    \centering
    \includegraphics[width=\columnwidth]{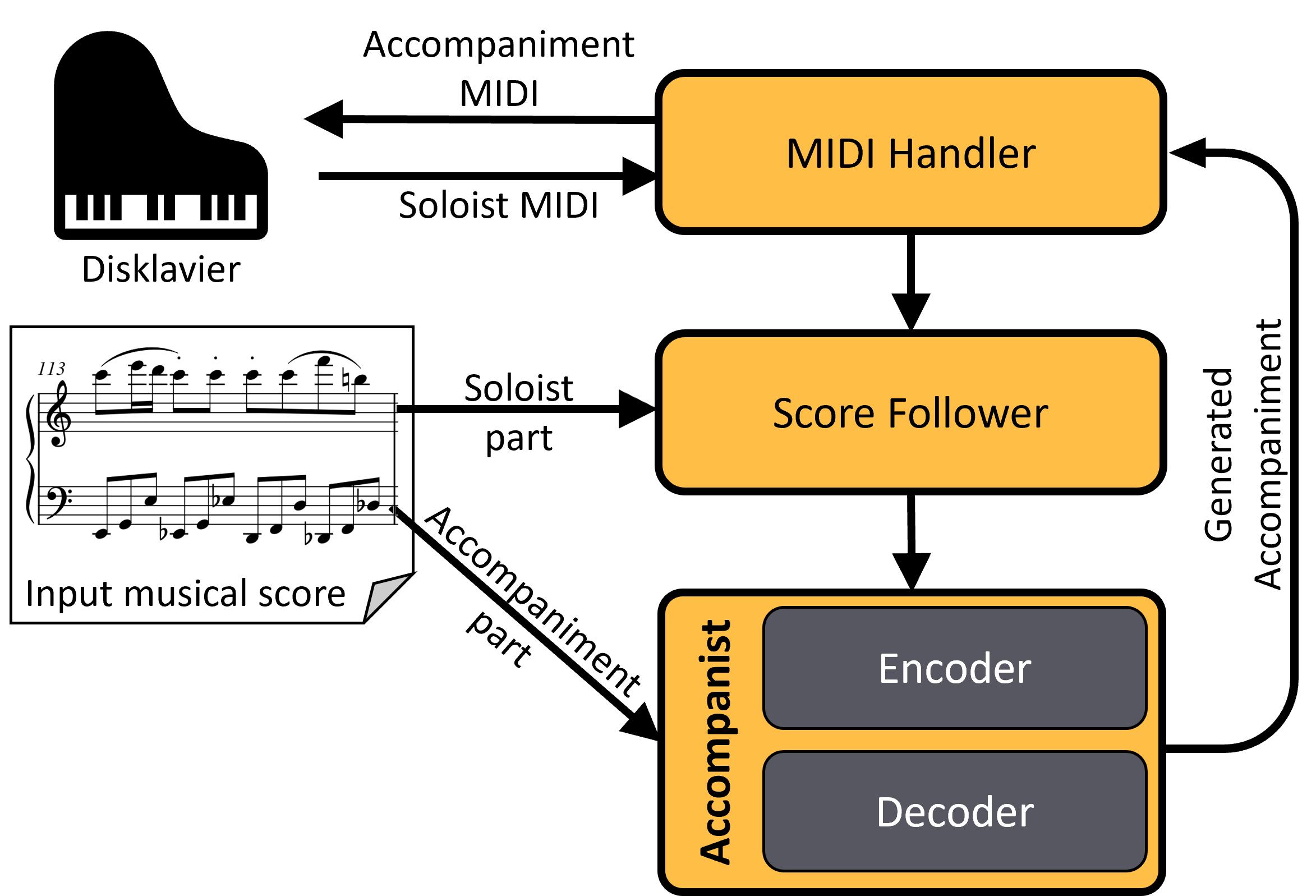}
    \caption{Modular architecture of the \Acc.}
    \label{fig:architecture}
\end{figure}

Let us consider a musical score with a solo and an accompaniment part, and a human player performing the solo part. 
The musical score is assumed to be in a digital symbolic format (e.g., MusicXML, MEI) and the soloist is performing on a MIDI-capable instrument. 
We use Partitura \cite{partitura_mec} and the Mido\footnote{\url{https://github.com/mido/mido/}} for handling scores and real-time MIDI input, respectively.
The \Acc generates a MIDI real-time expressive rendition of the accompaniment part 
that is conditioned on the tempo, dynamics and articulation of the soloist.

Figure~\ref{fig:architecture} presents a schematic view of the system.
Its main functionalities are implemented in two modules: the \texttt{Score Follower} and the \texttt{Accompanist}, which solve tasks 2 and 3 proposed by Dannenberg~\shortcite{dannenberg1984line}.
These modules are the focus of the rest of this section.
For practical implementation reasons, there is a third module, the \texttt{MIDI Handler}, which routes input and output MIDI messages. 
This module allows the \Acc to work in a variety of different configurations, including working directly with MIDI-capable instruments (e.g., MIDI controllers or player pianos) and being used with MIDI player software.

\begin{figure}[t]
    \centering
    \includegraphics[width=\columnwidth]{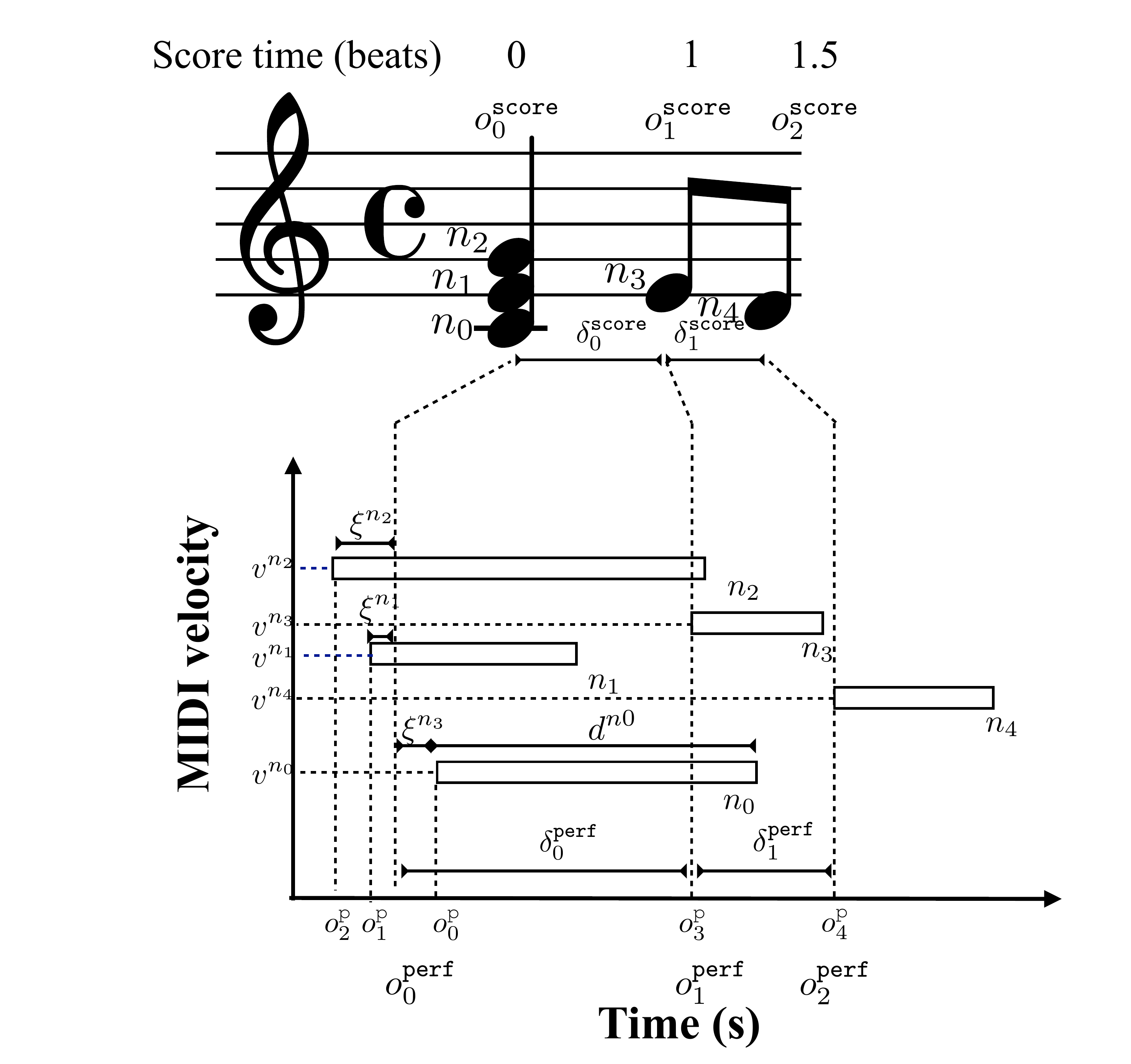}
    \caption{Excerpt of a MIDI performance (as a piano roll where the x-axis describes performance time in seconds and the y-axis is the MIDI velocity of the note) and its corresponding score, showcasing the elements for encoding/decoding an expressive performance.
    }
    \label{fig:performance_encoding}
\end{figure}

Before moving to a description of the \texttt{Score Follower} and \texttt{Accompanist} modules, we need to introduce some notation that is used in the rest of the paper.
Figure \ref{fig:performance_encoding} shows how  score and performance information is represented in the system.
We model the $i$-th  score note with a triple $(p^\texttt{s}_i, o^\texttt{s}_i,d^\texttt{s}_i)$ which corresponds to its MIDI pitch, score onset time and duration in musical beats.
Notes in a chord (i.e., notes are intended to be performed at the same time) will have the same score onset time, which we represent by $o^\texttt{score}$.
The time between two consecutive score onset times, the so-called inter-onset interval (IOI) is denoted by $\delta^{\texttt{score}}_i $.
The $i$-th performed note  can be described by the quadruple $(p^\texttt{p}_i,o^\texttt{p}_i,d^\texttt{p}_i,v^\texttt{p}_i)$, i.e., MIDI pitch, onset time and duration in seconds and MIDI velocity.  
We use $o^\texttt{perf}_i$ to denote the performed onset time corresponding to the notes at score onset time $o^\texttt{score}_i$.\footnote{Note that this onset time could be different than the onsets of the individual performed notes (as shown in Figure \ref{fig:performance_encoding}), since the notes in a chord are never played at exactly the same time.}
The performed IOI is denoted by $\delta^{\texttt{perf}}_i$. 

\subsection{Score Follower}\label{sec:score_follower}

Following our discussion in Section~\ref{sec:rel_score_foll}, we  implement two alternative score following approaches, one based on HMMs and the other using OLTW.
The score follower also contains a \textit{note tracker} object that keeps a record of the MIDI velocity $v^\texttt{p}$ and duration $d^\texttt{p}$ of the past performed notes.

In order to deal with polyphonic MIDI inputs, we use a windowing approach: incoming input MIDI messages are aggregated into non-overlapping 10\si{\milli\second} windows, and we consider that all of the played notes inside each of these windows correspond to the same position in the score (i.e., belong to the same chord) and have the same performed onset time $o^\texttt{perf}$ (the end of the window).
The input to the score follower is these  windows.
The full technical details of the models can be found in the Technical Appendix in the supplementary materials.

\subsubsection{HMM Follower}
The HMM-based score follower is based on the switching Kalman filter architecture, a hybrid probabilistic model which combines an HMM and a Kalman filter, whose parameters depend on the states of the HMM~\cite{Murphy:1998wd}. 
The observed variables of this model are the performed MIDI pitch  (the pitch of all of the notes inside the input 10\si{\milli\second} windows) and $\delta^\texttt{perf}$, the performed IOI.
The hidden variables are the set of score onset times $o^\texttt{score}$ plus some intermediate state for each score onset to account for possible insertions (modelled by the hidden states of the HMM), and the tempo of the performance (modelled by the Kalman filter part of the model).
We can then use the forward algorithm~\cite{rabiner1986introduction,Murphy:1998wd} to infer the position in the score in real-time.
The design of this score follower is partially based on the probabilistic score follower discussed in~\cite{Raphael2009OrchestralAF}.
While this model is very reliable for relatively simple sequential input, the HMM seems to struggle with complex music, particularly music that includes cross-rhythms, ornaments or large tempo changes (see e.g., the excerpt in Fig.~\ref{fig:tempo_model}).

\subsubsection{OLTW Follower}

To address the issues with the HMM-based score follower, we use OLTW, a causal dynamic programming algorithm for aligning sequences of different lengths which incrementally aligns in real-time a streamed sequence of unknown length (in our case, the input 10 \si{\milli\second} windows containing the MIDI performance of the soloist) to a known sequence, which we refer to as \emph{reference} (which represents the score)~\cite{dixon2005line}. 
 The performance of OLTW is controlled by 2 parameters, namely, a window size that controls how much context in the reference the algorithm considers at any given time, and a step size which controls the maximum step that OLTW is allowed to make at any given time. 
Following the research in real-time audio-based music alignment~\cite{Arzt2015RealTimeMT,Arzt2016thesis}, instead of aligning the performance directly to the score, we leverage the fact that human performers tend to play the same piece in a consistent way, and we align the input real-time performance to a previously recorded \emph{reference performance} (ideally by the same performer), which has been aligned to the score using offline alignment methods like the one by Nakamura et al.~\shortcite{nakamura2015real}.
Furthermore, we increase the robustness of the tracking by using an ensemble of OLTW score followers, each of which aligns the input performance to a different reference performance, and we aggregate the results by computing the mean position, as proposed by Arzt and Widmer~\shortcite{Arzt2015RealTimeMT}.

\subsection{Accompanist}\label{sec:accompanist}

The \texttt{Accompanist} module takes the score of the accompaniment part and uses the temporal alignments produced by the score follower, including the dynamics (MIDI velocity) and note duration to generate an expressive performance of the accompaniment in real-time.
This module consists of two submodules, an \texttt{Encoder} that translates the input performance into a set of \emph{expressive parameters} quantifying tempo, dynamics and articulation, and a \texttt{Decoder}, which takes these parameters to generate the performance of each of the notes in the accompaniment part.

\begin{figure}
    \centering
    \includegraphics[width=0.8\columnwidth]{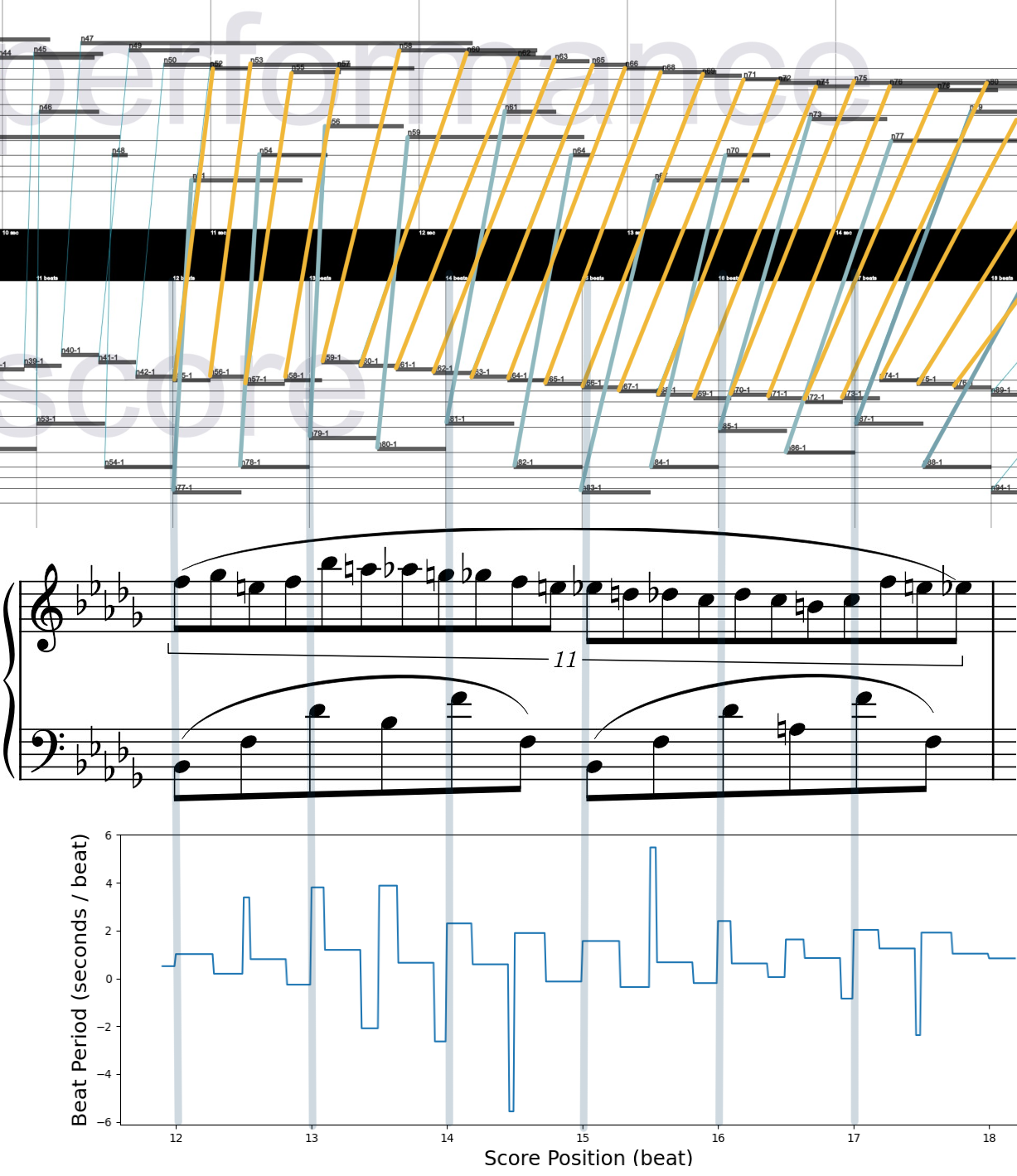}
    \caption{An excerpt of a performance of Chopin's Nocturne Op.~9 No.~1. The unusual cross-rhythm in the third measure (score in the middle) is usually played with substantial expressive freedom. Computing an IOI-based tempo based on an exact note alignment (piano rolls and connecting lines on top) leads to a jagged tempo curve (bottom), and even negative tempo values. This curve is unlikely to represent the more loose and ethereal subjective feel of the tempo in this passage.}
    \label{fig:tempo_model}
\end{figure}

Figure \ref{fig:performance_encoding} shows the different elements required to encode the performance of the soloist and decode the expressive accompaniment performance.
The performance of each of the notes can be encoded in 4 parameters: (1) \emph{MIDI velocity} $v_i^\texttt{p}$ for capturing expressive dynamics, (2) \emph{beat period}, defined as
 \begin{equation}
 b_i=\frac{\delta_i^{\texttt{perf}}}{\delta_i^{\texttt{score}}}\label{eq:beat_period}
 \end{equation} 
i.e., the duration of a beat in seconds (inversely proportional to tempo in beats per minute) which captures the local tempo deviations, (3) \emph{micro-timing deviations} (denoted by $\xi^i$ in Figure \ref{fig:performance_encoding}) which account for the onset time deviations of the notes from the global chord onset $o^\texttt{perf}$; and (4) the \emph{log articulation ratio} computed as
\begin{equation}
 a_i = \log_2 \frac{d_i^\texttt{p}}{d_i^\texttt{s} \cdot b_i}
 \end{equation} 
 which accounts for the ratio between performed note duration and notated duration at the current tempo).
 
The value of the beat period defined by Eq.~\ref{eq:beat_period} taken ``as is'' can be highly erratic in the case of real performances (see Figure~\ref{fig:tempo_model}), which typically contain large temporal fluctuations, and does not correspond with the human perception of tempo~\cite{Dixon:2006vi}, which results in very unnatural (and unmusical) performances.\footnote{We focus on tempo rather than musical meter, to determine when to trigger accompaniment events based on the soloist's real-time performance.} 
To address this issue, we add a tempo model inspired by the cognitive models of sensorimotor synchronization.
In particular, we tested three variants, namely a \emph{linear synchronization model} (L), which is a basic timekeeper model that adapts the tempo based on an error-correction mechanism based on the observed asynchrony~\cite{wing1973response,vorberg1996modeling}, a \emph{linear tempo expectation model} (LTE) which expands the linear model by incorporating tempo expectations extracted from the reference performances\footnote{Note that the LTE model implicitly captures musical meter.} (used for the OLTW score follower) and the \emph{joint adaptation anticipation model} (JADAM) proposed in \cite{van2013adaptation}, which includes both an error-correction term (the adaptation part) and a moving average estimate of the observed beat period (the anticipation part). 
Additionally, for our experiments, we use 3 baseline models that do not include asynchrony-based correction terms: a purely \emph{reactive model} (R) which uses Eq.~\ref{eq:beat_period} directly, a simple \emph{moving average} estimate of the beat period (MA) and a \emph{Kalman filter} (K).

\section{Quantitative Evaluation}\label{sec:evaluation}
We perform two experiments that focus on the quantitative evaluation of two interactive parts of our system. 
The first concerns the symbolic score follower, and the second evaluates the tempo models. 
Since, to the best of our knowledge, there are no available datasets of duet piano performances, we run both experiments using solo piano performances selected from  the Magaloff~\cite{magaloff} and Zeilinger~\cite{CancinoChacon:2017ht} datasets, which consist of performances recorded on computer controlled B\"osendorfer grand pianos which have been aligned to their corresponding scores.
The pieces are: Nocturnes Op.~9 Nos.~1 and 2, Etude Op.~10 No.~11, Nocturne Op.~15 No.~2, the Barcarole Op.~60 by F. Chopin, and the third movement of the Sonata Op.~53 (Waldstein) by L.~v.~Beethoven.
These pieces were selected because they contain interesting difficulties such as ornaments, trills and complex rhythms.

\subsection{Score Follower}
 
We evaluate the ability of our score follower to follow complex performances.
Since we only have one performance of each piece in the dataset, we artificially generate 5 reference performances per piece by adding zero-mean Gaussian noise with a standard deviation of 100\si{\milli\second} to the performed onset and offset times of the notes.
This process would be roughly similar (although less musical) to how musicians tend to play the same piece similarly across repeat performances \cite{demos2016flexibility}.
\begin{table}
    \centering
    \begin{tabular}{lrrrr}
\toprule
SF  &  Async  &  $\leq25\si{\milli\second}$ & $\leq50\si{\milli\second}$  &  $\leq100\si{\milli\second}$   \\
\toprule
HMM &645.7& 5.5& 5.5& 5.5\\
OLTW &60.6& 38.0& 63.3& 86.7\\
\bottomrule
\end{tabular}

    \caption{Alignment accuracy for the HMM- and OLTW-based score followers (SF).
    The first column represents the median absolute asynchrony in \si{\milli\second}, and the remaining columns represent the percentage of asynchrony which is less than 25ms, 50ms, and 100ms accordingly, of the same pieces presented in Table~\ref{tab:comparison_tuning_set}.
    }
    \label{tab:comparison_oltw_hmm}
\end{table}
Table \ref{tab:comparison_oltw_hmm} compares the alignment accuracy of the HMM and OLTW score followers, and Table \ref{tab:comparison_tuning_set} shows how well the OLTW works for the individual pieces. To put the values in perspective, between two human instrument performers, average asynchrony varies between 30-40ms~\cite{keller2007pianists} but also depends on the difficulty of the piece being played.
These results show how OLTW can leverage information from prerecorded reference performances to improve the alignment of complex pieces (other than the Waldstein Sonata, which is arguably one of the most difficult pieces in the standard classical piano repertoire and more challenging than the other pieces in the dataset), while the HMM struggles using only information in the score (which does not include, for example, the ornament notes like trills). We also notice that the distribution of the HMM performance is bimodal, i.e., either very low or very high asynchrony.

\begin{table}[t]
    \centering
    
\small
\begin{tabular}{lrrrr}
\toprule
Piece  &  Async  &  $\leq25\si{\milli\second}$ & $\leq50\si{\milli\second}$  &  $\leq100\si{\milli\second}$   \\
\toprule
B. Op.~53 3rd.~mov.& 1,813 & 18.4 & 29.3 & 44.4 \\
C. Op.~9 No.~1 & 44 & 59.1 & 75.5 & 90.9 \\
C. Op.~9 No.~2 & 59 & 42.3 & 62.7 & 87.0 \\
C. Op.~10 No.~11 & 63 & 33.6 & 63.9 & 91.3\\
C. Op.~60 & 106 & 20.0 & 36.8 & 65.1\\
\bottomrule
\end{tabular}
    \caption{Alignment accuracy of OLTW score follower per piece. Column 2 presents the mean asynchrony of the score followers in milliseconds. Columns 3, 4, 5 present the percentage of asynchrony which is less than 25ms, 50ms and 100ms accordingly. 
    Hyperparameters for the OLTW score follower are 2\si{\second} window size and 0.1\si{\second} of step size.}
    \label{tab:comparison_tuning_set}
\end{table}

\subsection{Tempo Model}
\label{sec:tempo_models}

In this experiment, we evaluate the tempo models we describe above.
Specifically, for each incoming onset in the soloist performance, and given the tempo computed at the previous step, the tempo model predicts a new tempo value and the next soloist onset.
We then compare this prediction with the ground truth tempo and onset in the soloist performance.
We run a grid search over roughly 800 parameter combinations and report values for the best parameters for each model. In case of disagreement between tempo and onset, we select the best parameters based on onset prediction.
Table \ref{tab:table_tempo} shows the results of this experiment.

\begin{table}[t]
    \centering

\begin{tabular}{lrr}
\toprule
Method       & Onset Error (ms)                & Tempo Error (ms/beat)   \\
\toprule
R      & 4,279.7               & 209.3 \\
MA       & 4,271.0              & 203.5 \\
L       & 81.9              & 173.1 \\
LTE       & 23.3              & 63.3 \\
JADAM       & 94.1              & 190.4 \\
KT       & 1,154.2              & 177.3 \\
\bottomrule
\end{tabular}

    \caption{Average absolute errors in both tempo (milliseconds / beat) and onset (milliseconds) prediction for each tempo model.} 
    \label{tab:table_tempo}
\end{table}

The performance of the LTE model can be explained by its design advantage: this model makes use of a reference performance of the same piece to guide its estimation. In normal usage of the \Acc, this would be another performance the soloist recorded during their rehearsal with the system. In this experiment, this is obtained by adding Gaussian noise to the onsets of the soloist performance we consider.
If the reference performance is very close to the test performance, which is the case in our experiment,
the model prediction is extremely accurate. If the reference is missing, the model falls back to the linear model.

Note that a high predictive accuracy on solo piano performances does not necessarily reflect the actual goal of these models in our accompaniment setting. In our usage, the models serve two purposes: to predict the next onset and to smooth the precise, but jagged and unmusical tempo estimation based on the score follower output (see Figure \ref{fig:tempo_model}). In this experiment, we only evaluate the first purpose, but a very high accuracy might even indicate an unsuitable algorithm that fails to meaningfully produce the human perception of tempo and follows the input too tightly.
In the next section, we discuss this issue from the performer's perspective.

\section{Usage and Discussion}\label{sec:discussion}

In this section, we detail the different situations in which the \Acc was tested and the feedback that this generated. We then start a critical discussion about music co-performance and the challenges of human-machine collaboration in this context.

\subsection{Public Live Demonstration}
The \Acc was presented at several public venues, both to scientific and non-scientific audiences,
in a two-part format. 
The first part consisted of a co-performance of Brahms' 5th Hungarian Dance (for four hands) by one of the authors. 
The system settings were manually chosen based on the preferences of the musician. 
In the second part, we invited people from the public to play with the system. 
To make the experience pleasant for people with different musical expertise without any preparation time, we selected simple pieces, in which the player only plays a short monophonic melody with the \Acc taking care of the accompaniment (left hand, mostly). 

\subsection{Musician Feedback}

\textit{In lieu} of a systematic user study with a sizeable number of different pianists and pieces, and as a first step towards getting a qualitative understanding of the most pressing problems that need to be addressed next,
we here reproduce personal `testimonials' by three of the authors of this paper (two of them professionally trained pianists), who have the most experience with playing with the \Acc. They worked on three pieces: Rondo in A major D.951 by F.~Schubert, Piano Sonata K381 by W.A.~Mozart, and Hungarian Dance No.~5 by J.~Brahms.
In Section~\ref{sec:machine-human_collaboration} we will then see what insights we can distil from these personal reports.

\paragraph{Musician 1.}
Playing with the \Acc has been an enlightening experience.
As both the main developer of the system and the person that has played with it the most, I am very aware of its shortcomings, but I am also very pleased (and perhaps, dare I say proud) of each (small) breakthrough.
Playing with the system has broadened my perspectives on what aspects are important for ensemble performance, in particular 
it has made me realize the importance of understanding and communicating shared expressive intentions between musical partners.
Every time that I play duets with other humans, I find myself thinking of what the \Acc would do, and it has made me more aware of how difficult it is to play with someone else, and how easily can things go wrong when the expressive intentions for the piece have not been discussed (e.g., when sight-reading unfamiliar music).
While the \Acc is still (very) far from being able to accompany a performance of Rachmaninoff's 3rd Piano Concerto (my personal goal for the system), it can be a rewarding experience.
For example, in the performance of Brahms' piece\footnote{\url{https://youtu.be/Wtxcqp-sQ\_4}}, it is very nice how the system slows down at the fermata at the end of part B, and then just starts playing on time afterwards.
In those cases, I can (almost) forget that I'm playing with an artificial partner and just focus on the music that I'm playing.

\paragraph{Musician 2.}
Playing with the \Acc was a fun, albeit somewhat troublesome experience. My piece starts with an eighth-note chord, after which I start playing a sequence of sixteenth notes, to which the system then responds with a similar sequence before we play along jointly. In the initial trials, the system would run away immediately by starting off with a crazy fast tempo. Even if I jumped in at the correct position (i.e., current system score position), it was not possible anymore to play jointly together, as the accompaniment would not slow down sufficiently.
After a lot of trial and error, we eventually figured out this seemingly random behaviour was caused by the initial chord and the (what the \Acc perceived as) ``pause'' before the sixteenth notes sequence, which caused the score follower to frantically align notes where there were none, and the tempo model to adjust the global tempo accordingly. After we cut this initial chord (i.e., starting the piece at my sequence directly), we were playing jointly and in the same (global) tempo, and the \Acc then was able to respond effortlessly to my playing style, especially in terms of (micro-)timing and dynamics. Overall, I had the impression that the system was \emph{re}acting more than it was acting on its own, similar to when rehearsing for the first time with a human ensemble partner who is not yet too familiar with ensemble playing.

\paragraph{Musician 3.}
Playing with the \Acc was an encounter of the third kind, and
a rather stressful one, for me. The moment that epitomizes it all is the
very beginning of a performance: you play the first note, and you just
\textit{hope} that the \Acc will join in, in the right tempo. It
(almost) always does, in the end, but the very fact that you have to
worry about it points to the central problem: a complete lack of natural
communication and trust. For the next few bars, you are busy being
relieved that the start `worked', and then, for the rest of the piece,
you very consciously focus on pulling it along, tricking it into
speeding up or slowing down (which might or might not be necessary; but
you just don't trust it), probing how much spontaneous change it can
handle -- in short: you are constantly focused on \textit{it}. The
result of my efforts can be seen in this video\footnote{\url{https://youtu.be/qEocywdruco}} (with apologies
to Franz Schubert) -- my playing is unrelaxed, unnatural throughout.
So: this research teaches us as much about what's missing as it does
about probabilistic tempo modeling or performance prediction. To me, the
ultimate basis is \textit{trust}: in each other's musical understanding,
but also in my partner's ability and readiness to sense when I struggle
and support me.

\subsection{On Human-Machine Expressive Collaboration}
\label{sec:machine-human_collaboration}

Using the framework of togetherness proposed by Bishop \shortcite{bishop2022focus}, the extent of expressive musical collaboration can be placed on a spectrum ranging from mere awareness of others
(i.e., the lower bound of interaction) to the experiential process of cognitive and emotional alignment.
While collaborative musical interaction between humans can be studied from this phenomenological standpoint, the same approach cannot be taken when investigating human-machine collaboration, as systems inherently do not possess any internal state and are not capable of perceiving music (or sound, for that matter) other than in the way they are designed and built to do.
Indeed, our musical experiences with the system suggest that the 
interaction the \Acc has with the soloist is biased towards following and, for the most part, our system misses out on making its own decision.
For example, we observe that when the system ``believes'' that the human player is slowing down (which may not be the case), it slows down in response, which causes the human player to slow down as well, which in turn causes the system to slow down even more. The system reaction, in this case, is contrary to what a human accompanist would do, i.e., to try to keep the tempo stable, and results in a decreased sense of togetherness.

In Section~\ref{sec:rel_work} we reported the subtasks existing accompaniment systems focus on: note detection (for audio systems), score following, and expressive accompaniment generation.
We believe that a missing fourth point should be added to this list: \emph{modeling the feedback loop with the human partner}.
This means that
the system needs to ``understand'' when to be more \textit{reactive} and follow the human player, and when more \textit{proactive} and lead the performance.
Some system upgrades that would help in this direction are: further study of tempo models, the usage of the long-term musical structure of the piece, and the introduction of visual cues. The latter can be considered in the music co-performing context, as useful as non-verbal communication is in the speech context. For example, let us consider the challenging situation of a chord after a long rest. Musicians would strongly rely on visual cues, such as breathing, nod, and hand gestures, to synchronize. Given our previous considerations on leading and following, we would need visual cues for both the system (e.g., signals from body sensors or a camera), and the player (e.g., video or haptic feedback on the state of the tempo model).

\section{Conclusion}\label{sec:conclusion}

This paper introduces the \Acc, an automatic system capable of accompanying a human soloist performing a given musical score.
It can work with a variety of different inputs, as long as the full musical score is given in a symbolic format, the soloist performance can be encoded in MIDI data, and a system that translates MIDI outputs to sound is available.
We describe the two tasks our accompanist must perform: online score following and automatic accompaniment generation, and we approach their interaction with cognitively plausible models inspired by research on sensorimotor synchronization.
We perform two quantitative experiments to evaluate critical parts of our system, test it at multiple public live demonstration, and evaluate it personally from a musician's perspective.
Based on the experimental results, audience feedback and our own experience, we start a critical discussion on the still open challenges for accompaniment systems to be precise and reliable musical partners

Future development of the \Acc will go in the direction of modeling the feedback loop with the human partner, for the system to understand when to follow the soloist and when to lead the performance.
Furthermore, we plan a systematic user evaluation of the entire system in different configurations, based e.g., in the work by Zhou et al.~\shortcite{Zhou2023}.
Moreover, we are working on a new end-to-end accompanist part, that would produce an expressive accompaniment conditioned to the soloist performance in a single step, with the advantage of sharing the information about the two tasks, to improve both.
Finally, we plan on introducing visual cues for both the soloist, in the form of visual feedback on the computed tempo, and the accompaniment system, in the form of signals from body sensors or a camera.

\appendix

\section*{Ethical Statement}

There are no ethical issues.

\section*{Acknowledgments}
We want to thank Laura Bishop for her contributing recordings to train the ACCompanion.
This work is supported by the European Research Council (ERC) under the EU's Horizon 2020 research \& innovation programme, grant agreement No. 10101937 ("Wither Music?").

\bibliographystyle{named}
\bibliography{ijcai23}

\newpage

\section{Technical Appendix}\label{sec:technical_appendix}

This Appendix follows the notation introduced in Section \ref{sec:architecture} (in particular Figure \ref{fig:performance_encoding}), but we drop the superscript $\texttt{p}$ of performed onsets to unclutter notation.
In the following description of the tempo models, we denote the observed performed onset time of the $n$-th onset in the score as $o_n$ and the onset time predicted by the synchronization models as $\hat{o}_n$.
The asynchrony between these onsets is denoted as $A_n = \hat{o}_n - o_n$ and the observed beat period is given as 
$\tau_n = \frac{\delta_n^{\texttt{perf}}}{\delta_n^{\texttt{score}}}$, with $\tau_0$ being the initial tempo set by the performer as a hyper parameter.

\subsection{Reactive Sync Model (R)}\mbox{} \\
The next accompaniment onset is ``estimated'' at the same time as the last observed performed onset, making this model purely reactive.
\begin{align}
\hat{o}_{n+1} & = o_n + b_n \delta^{\texttt{score}}_n\label{eq:rsm_eo}\\
b_{n+1} &= \tau_n
\end{align}

\subsection{Moving Average Sync Model (MA)}\mbox{} \\
This model is very similar to the first baseline, except for its estimation of the global tempo, which uses a weighted average of the last and current tempo estimate. 
\begin{align}
\hat{o}_{n+1} & = o_n + b_n \delta^{\texttt{score}}_n\label{eq:masm_eo}\\
b_{n+1} & = \eta_{\text{MA}} b_n  + (1 - \eta_{\text{MA}} ) \tau_n\label{eq:masm_bp}
\end{align}
\noindent where $\eta_{\text{MA}}$ is a constant parameter.

\subsection{Linear SMS Model (L)}\mbox{} \\
Onset and beat period estimates ${o_{n+1}}$ and ${b_{n+1}}$ are computed as follows:

\begin{align}
\hat{o}_{n+1} & = \hat{o}_n + b_n \delta^{\texttt{score}}_n - \eta^{o} A_n \\
b_{n+1} & = \begin{cases}b_{n} - \eta^{b} A_n, & \text{if } A_n<0, \\b_{n} - 2\eta^{b}A_n, & \text{else} \end{cases} \label{eq:lsm_bp}
\end{align}

where ${\eta^{o}}$ and ${\eta^{b}}$ are the learning rates for the onset and beat period, respectively.

\subsection{Linear Tempo Expectation  Model (LTE)}\mbox{} \\
This model is similar to the previous one, but includes an anticipation (expectation) term in the computation of the beat period: 
\begin{align}
\hat{o}_{n+1} & = \hat{o}_n + b_n \delta^{\texttt{score}}_n - \eta^{o} A_n \\
b_{n+1} & = \phi(o^{\texttt{score}}_{n+1}) - \eta^{b} A_n, 
\end{align}

\noindent where $\phi(o^{\texttt{score}}_{n})$ is a function that computes an estimate of the beat period at score onset time $o^{\texttt{score}}_{n}$ based on the tempo of the reference performance(s) and ${\eta^{o}}$ and ${\eta^{b}}$ are constant parameters that represent the learning rates for the onset and beat period, respectively.

\subsection{Joint Adaptation Anticipation Model (JADAM)}

This model includes both an error-correction term (the adaptation part) and a moving average estimate of the observed beat period (the anticipation part).
\begin{enumerate}
\item Adaptation Module
\begin{align}
\hat{o}^{\text{ad}}_{n+1} & = \hat{o}_n + b_n \delta^{\texttt{score}}_n - \eta^{o}_{\text{J}} A_n \\
b_{n+1} & = b_{n} - \eta^{b}_{\text{J}}  A_n
\end{align}

\item Anticipation Module
\begin{align}
\hat{\tau}_n &= \eta_{\text{J}}^{b}(2 \tau_n - \tau_{n-1}) + (1- \eta_{\text{J}}^{b}) \tau_n \\
\hat{o}^{\text{an}}_{n+1} & = o_n + \hat{\tau}_n \delta^{\texttt{score}}_n
\end{align}

\item Joint Module
\begin{align}
\hat{A}_n &= \hat{o}^{\text{ad}}_{n+1} - \hat{o}^{\text{an}}_{n+1} \\
\hat{o}_{n+1} &= \hat{o}^{\text{an}}_{n+1} - \eta_{\text{J}}^{a} \hat{A}_n
\end{align}
\end{enumerate}

Parameters ${\eta_\text{J}^{o}}$ and ${\eta_\text{J}^{a}}$ and  ${\eta_\text{J}^{b}}$ are learning rates for the prediction of the onset of the adaptation and joint modules and the learning rate for the beat period, respectively.

\subsection{Kalman Tempo  Model (KT)}

The observed variable of the Kalman filter is the performed onset time, and the beat period is the latent variable.
The updates are computed as follows:
\begin{align}
\hat{b}_n &= \alpha_{\text{K}} b_n \\
v_n &= \gamma_{\text{K}}^2 \hat{v}_n + \beta_{\text{K}} \\
\hat{A}_n &= \delta^{\texttt{perf}}_n - \hat{b}_n \delta^{\texttt{score}}_n \\
\kappa_n &= \frac{v_n \delta^{\texttt{score}}_n}{v_n\delta^{\texttt{score}^2}_n + \lambda_{\text{K}}}\\
b_{n+1} &= \hat{b}_n + \kappa_n \hat{A}_n \\
\hat{v}_{n+1} &= (1 - \kappa_n \delta^{\texttt{score}}_n) v_n \\
\hat{o}_{n + 1} &= \hat{o}_n + b_{n+1} \delta^{\texttt{score}}_n
\end{align}
where $\alpha_\text{K}$, $\beta_\text{K}$, $\gamma_\text{K}$ and $\lambda_\text{K}$ are the parameters of the model.

\end{document}